\documentstyle[epsfig,psfig]{mn2e}

\title{The Submillimetre Properties of Ultraluminous Infrared Galaxies}
\author[D.L. Clements et al.]
        {D.L. Clements$^1$, L. Dunne$^2$, S.Eales$^3$\\
        $^1$Imperial College, London, Blackett Lab, Prince Consort Road, London SW7 2AZ, UK\\
        $^2$Centre for Astronomy \& Particle Theory, School of Physics \& Astronomy, University of Nottingham, University Park, Nottingham NG7 2RD\\
        $^3$School of Physics and Astronomy, Cardiff University, 5 The Parade, Cardiff, Wales CF24 3YB\\
\\}
\date{}

\pagerange{\pageref{firstpage}--\pageref{lastpage}}
\pubyear{}

\begin{document}

\maketitle

\label{firstpage}

\begin{abstract}
We present the results of SCUBA observations of a complete sample of local ULIRGs. Twenty of the initial sample of 23 sources are detected at 850$\mu$m and nearly half of the objects are also detected at 450$\mu$m. This data is combined with existing observations of a further seven ULIRGs to produce the largest sample of submm observations of ULIRGs currently available. We use similar techniques to the SLUGS survey to fit dust spectral energy distributions (SEDs) to their far-IR emission. We find that ULIRGs have a higher dust temperature than lower luminosity objects (42K compared to 35K) and a steeper emissivity index. For those objects where 450$\mu$m fluxes are available we also attempt a two component dust SED fit, with warm and cool dust and a dust emissivity index of $\beta$=2. Such a model has been found to be a good fit to lower luminosity systems. We find that it also works well for ULIRGs, but that ULIRGs have a smaller cold dust component. Comparison of the dust mass derived for ULIRGs and more normal spiral galaxies suggests that the dust content of a ULIRG is simply the combined dust content of the two galaxies whose merger has triggered the ULIRG activity. We examine the high end of the 850$\mu$m luminosity function and find results consistent with those of the earlier SLUGS survey. We also find that ULIRGs make up only about 50\% of the high end of the 850$\mu$m luminosity function, with LIRGs containing a large mass of cool dust likely to be responsible for the rest.
\end{abstract}


\section{Introduction}

Ultraluminous Infrared Galaxies (ULIRGs) are exceptional local universe objects with far-IR (L$_{1-1000\mu m}$) luminosities in excess of 10$^{12} L_{\odot}$ and where the bulk of the luminosity, $\sim$99\%, is emitted in the far-infrared
(Wright et al., 1984; Sanders et al., 1988; Sanders \& Mirbel 1996 and references therein). Optical and near-IR imaging (eg. Clements et al., 1996; Murphy et al., 1996) has shown that they are almost universally galaxy mergers while spectral energy distribution (SED) analysis (eg. Farrah et al., 2003) reveals that they are largely powered by starbursts. Their role in galaxy evolution is unclear but it seems likely that they are an intermediate starbursting phase that takes place when a major merger converts two colliding, gas- and dust-rich spiral galaxies into an elliptical galaxy (Barnes \& Hernquist 1991). Such a merger event may also be associated with the growth of a supermassive black hole (SMBH) at the centre of the merging galaxy (eg. Kauffmann \& Haehnelt, 2002) and resulting AGN activity. Indeed, ULIRGs have been suggested as the sites of future quasar activity once the AGN, obscured by the large mass of gas and dust driven to the galactic centre by the merger, has managed to emerge. These issues have been studied by Dasyra et al. (2006a, 2006b) through observations of stellar kinematics in ULIRGs. They confirm that ULIRGs are major mergers that result in the formation of elliptical galaxies and the likely fueling of a 10$^7$-10$^8 M_{\odot}$ central black hole.  ULIRGs are also thought to be the local counterparts of the submm luminous galaxies uncovered at large redshifts by surveys using the current generation of submm array camera, such as SCUBA and MAMBO (Smail et al., 1997; Eales et al., 2000; Eales et al., 2003), though SMGs and local ULIRGs appear to lie in somewhat different parts of the temperature-luminosity plane (Yang et al., 2007). Such objects are thought to be significant contributors to the Cosmic Infrared Background (Puget et al., 1996; Fixsen et al., 1998). The study of local ULIRGs and their relationship to the broader population of galaxies, can thus illuminate not only sites of extreme star formation and AGN triggering in the local universe, but may also have implications for the formation of galaxies and quasars in much earlier stages of the universe when such objects were much more common.

Our knowledge of submm emission from galaxies is currently limited to studies of only a few hundred galaxies (eg. Dunne et al., 2000; Dunne \& Eales, 2001; Vlahakis et al., 2005; Lisenfeld et al., 2000) largely selected from flux limited samples from IRAS or the optical. These studies set the best baseline for our knowledge of local galaxies in the submm until more sensitive instruments such as SCUBA2 or data at other far-IR/submm wavelengths inaccessible from the ground become available from Herschel. These studies have determined the local submm (850$\mu$m) luminosity function and the dust mass function and have uncovered evidence that significant masses of cold dust (T$<$30K), undetectable by IRAS, appear to be present in most quiescent spiral galaxies, a result recently confirmed by Spitzer (Willmer et al., 2009). Our knowledge of the submm SEDs of ULIRGs, in contrast, in relatively poor, with fewer than 15 local ULIRGs studied at these wavelengths so far (eg. Farrah et al., 2003), none of which have been selected as part of a statistically complete sample. This is largely because local ULIRGs remain very faint at submm wavelengths compared to the submm sensitivities currently available, so a significant amount of observing time in good conditions must be dedicated to any coherent survey of their submm properties. We here report such a project, aimed at obtaining 850$\mu$m and, where possible, 450$\mu$m, fluxes for a complete sample of ULIRGs selected from the 1Jy ULIRG survey (Kim et al., 2002). We combine this statistically complete sample with existing data from the literature to produce the largest set of SCUBA submm continuum data on local ULIRGs that is currently available. This will provide a stepping stone for future work on ULIRGs with SCUBA2 and Herschel. One comparable data set available at shorter wavelengths (350$\mu$m) and somewhat higher redshift than the present work is that of Yang et al. (2007), who used SHARCII to observe a sample of 36 ULIRGs identified through a cross matching of IRAS Faint Source Survey (FSC) objects and radio sources from the Faint Images of the Radio Sky at Twenty cm (FIRST) survey. This sample consisted of 36 sources of which 28 were detected at S/N$>3$ so the two samples are comparable in size.

The paper is organised as follows. We first discuss the sample of objects selected and the additional sources from the literature (section 2). We then detail the observations undertaken and present the fluxes obtained for our targets, together with literature data for those already observed and for the additional sources outside the complete sample (Section 3). We then examine the dust SEDs of our ULIRGs, using both one and two component dust models (Section 4), followed by an examination of the contribution of ULIRGs to the local submm luminosity function (Section 5). We then draw our conclusions (Section 6). Throughout the paper we assume H$_0 = 75 kms^{-1}Mpc^{-1}$.
%, $\Omega_m = 0.3$ and $\Omega_{\Lambda} = 0.7$ throughout.

\section{The ULIRG Sample}

The initial target list comprises a complete sample of local ULIRGs selected on the basis of the IRAS 1Jy survey of Kim et al. (2002) with updated photometry from NED. The selection criteria are that the source should have a 60$\mu$m flux $>$1.5 Jy and have 8h $<$ RA $<$ 17.5h, and 0$<$ DEC $<$ 60. [The B1950 coordinates of the 1Jy catalog are used for the selection.] The survey thus covers 1.55 steradians on the sky. 

This selection from the 1Jy ULIRG survey finds 23 ULIRGs in this area. Updated photometry from NED eliminates one source, as its new flux drops below the 1.5Jy limit (IRAS 13443+0802). 
An additional ULIRG, 15250+3609 is included which is missing from the 1Jy survey. The original 1Jy survey ULIRG classification was based on a $q_0 = 1$ cosmology. Two additional sources would be included as ULIRGs in the 1Jy survey if $q_0 = 0$, 10190+1322 and 10565+2448. 

To this complete sample we add an additional seven ULIRGs from the literature which are detected at 850$\mu$m. These include IRAS 10565+2448, as well as famous ULIRGs such as NGC6240 and IRAS 22491+1808 which do not fit the selection constraints of the main sample. Taken together we believe these thirty objects are the largest sample of ULIRGs with (in all but two cases) 850$\mu$m detections currently available.

\subsection{Non-Thermal Sources in the Main Sample}

Two sources in the complete sample are dominated by non-thermal emission, 3C273 (12265+0219) and 4C+12.5 (13451+1222). Data for these came from archival sources. They are both well studied radio sources where much of the 
submillimetre output is non-thermal in nature. Because of this they are discounted in our  analysis of the dust properties of ULIRGs. However, it should be noted that these two non-thermal sources represent about 10\% of the complete ULIRG sample. One might thus expect non-thermal sources to contribute at this level to any deeper flux limited sample of ULIRGs selected, for example from surveys conducted using Spitzer or Herschel. It is also interesting to note that the fraction of non-thermal dominated sources is similar to the fraction of radio-loud objects in the quasar population. These two sources are not included in our consideration of the 850$\mu$m luminosity function but if they were would lie well beyond the highest luminosity bin for the thermal sources.

\section{Observations and Data Reduction}

The majority of the sources in our complete sample had not previously been observed in the submm.
We thus conducted observations using the SCUBA bolometer array camera at the JCMT (Holland et al. 1999) to obtain fluxes at 850 and, when weather conditions permitted, 450$\mu$m. The data was obtained over several years as part of the JCMT queue observing system. The dates of observations are thus spread over a considerable time, details of which are given in Table \ref{table1}. The observations were taken in standard chopped photometry mode and reduced in the usual manner using the SURF data reduction software (Jenness \& Lightfoot, 2000). Weather conditions ranged from Grade 3 to Grade 1 for our observations, with 450$\mu$m data being useful in only the best weather.

Flux calibration was achieved using a range of primary and secondary JCMT calibrators, including the planets Uranus and Mars and the secondary calibrators CRL 618, OH231.8 and CRL2688. Where data on a single object were obtained on more than one night, the calibrated fluxes were combined using normal $\sigma$ weighted methods. The results of our observations are included in Table \ref{table2}, together with fluxes obtained from the literature and other basic information on the objects.

Of our complete sample of 23 ULIRGs 850$\mu$m fluxes were detected in 21. A limit was obtained for one (12018+1941) and one source was not observed (16487+5447). The sample is thus 96\% observationally complete and 91\% complete in terms of recovered fluxes. New 450$\mu$m fluxes were obtained for 5 sources which were observed in the best conditions. When archival data, listed in Table \ref{table3}, is added we have 450$\mu$m fluxes for ten sources in the complete sample, ie. 43\% of sources.

\subsection{Correcting for Extended Emission}

The JCMT beam at 850$\mu$m and 450$\mu$m is 15 and 7 arcsec respectively so there is a possibility that some extended emission might be missed from our chopped observations. We examined the results of K-band imaging of these objects by Kim et al. (2002) to determine the likely effects on our photometry. In all cases the JCMT was pointed at the bright optical and K band source identified with the ULIRG by Kim et al. (2002). Most of our sources are found to have compact nuclear regions.  Arp220, the nearest ULIRG and the prototype for the whole class, is found to have 850$\mu$m emission extended by less than 2 arcsec, ie. 1kpc at its redshift of 0.018 (Sakamoto et al., 2008). We thus deem it unlikely that our other sources with compact nuclei will have submm emission extended beyond our beams. Two sources (13539+2920 \& 16474+3430) have double nuclei separated by 4.2 and 3.5 arcseconds respectively. This is much smaller than the 850$\mu$m beam used to collect our data so there should be no issues with their fluxes (no useful fluxes were obtained at 450$\mu$m for these objects so the smaller beam in that band is not an issue). 16487+5447 also has two nuclei separated by 3.1 arcsec but no submm observations were obtained for this source. Three other sources have components separated on larger scales. 14349+5332 has a weak component, a factor of 8 fainter at K than the main source, separated by 29 arcsec, 15001+1432 has a second component 43 times fainter at K separated by 20.1 arcsec, while 17028+5817 has a second component separated by 13 arcsec that is 3.2 times fainter at K. Any 850$\mu$m flux from these secondary nuclei will fall outside our beam and will thus not have been accounted for by our observations. To correct for this we assume that these fainter separated components will make an additional submm flux contribution that matches their contribution to the K band light.  We add this additional flux to that observed by the JCMT. Because of the uncertainties in this estimation we also increase the errors in the flux determined for these objects by adding this additional flux as a systematic error to the total flux. 

For sources from the literature the most nearby objects in the sample have fluxes taken from Dunne \& Eales (2001) and so are covered by mapping observations. Extended emission in these objects will be fully accounted for. Klaas et al. (2001) examined NGC6240 and 17208-0014 with jiggle maps and found no evidence for emission extended beyond the beam. Extended emission in the rest of their sample was concluded to be insignificant given the flux uncertainties.

\begin{table*}
\label{table1}
\begin{tabular}{ccc}\hline
Target&Observation Date&Comments\\ \hline
10378+1108&14 Dec. 2002, 12 Jan. 2004&\\
10494+4424&19 Sept. 2002 \& 4 Jan. 2003&Second night Grade 1\\
11119+3257&18 Jan. 2003&\\
11506+1331&4 Jan. 2003, 13 Jan. 04&Both Grade 1\\
12018+1941&27 Jan. 2003, 12 Jan. 2004&Second night Grade 1\\
13442+2321&4 Jan. 2003&Grade 1\\
13451+1232&17 Mar. 1999&Grade 1, data from archive\\
13510+0442&19 Jan. 2003&\\
13539+2920&28 Jan. 2003&\\
14060+2919&17 Jan. 2003&\\
14394+5332&12 Jan. 2004&Grade 1\\
15001+1432&11 Jan. 2003&Grade 1\\
15206+3342&4 Jan. 2004&\\
16474+3430&18 Sept. 2003&\\
17028+5817&6 Aug 2001&\\
\end{tabular}
\caption{Observing Log. 
Observation dates for the sources observed for this paper. Data on other sources comes from the literature.}
\end{table*}

\begin{table*}
\label{table2}
\begin{tabular}{cccccc}\hline
Source&F60 (mJy)&F100 (mJy)&F450 (mJy)& F850 (mJy)&   Redshift\\ \hline
08572+3915&     7433$\pm$ 372&  4588$\pm$ 367&&                 17.0$\pm$ 7.0$^2$&      0.05835\\
10378+1108&  2281$\pm$ 137&  1816$\pm$ 163&&                 4.4$\pm$ 1.1&           0.1362\\
10494+4424&  3527$\pm$ 212&  5412$\pm$ 325&134.0$\pm$ 30.0&16.0$\pm$2.3&    0.09206\\
11119+3257&  1588$\pm$ 175&  1523$\pm$ 168&&                 5.6$\pm$ 1.9&           0.1890\\
%11506+1331&  2583$\pm$ 155&  3323$\pm$ 266&  35.0$\pm$ 11&  4.3$\pm$ 1.1&           0.12729\\
11506+1331&  2583$\pm$ 155&  3323$\pm$ 266&  &  4.3$\pm$ 1.1&           0.12729\\
12018+1941&  1761$\pm$ 123&  1776$\pm$ 231&&                 0.0$\pm$ 1.4$^{*}$&           0.1686\\
12112+0305&8503$\pm$ 510&9976$\pm$ 800&483$\pm$136$^1$&  49.0$\pm$ 10.0$^2$&     0.07332\\
12265+0219 (3C273)&  2000$\pm$ 140&     2891$\pm$ 203&&         6454.0$\pm$ 649.0$^3$&   0.1583\\
12540+5708 (Mrk231)&31990$\pm$ 1600&30290$\pm$ 1211&513.0$\pm$ 169.0$^4$&       96.0$\pm$ 18.0$^4$&     0.04217\\
13428+5608 (Mrk273)&21740$\pm$ 870&     21380$\pm$ 860& 707.0$\pm$ 245.0$^4$&   84.0$\pm$ 22.0$^4$&     0.03778\\
13442+2321&  1625$\pm$ 146&  2259$\pm$ 181&  107.0$\pm$ 25&        15.5$\pm$ 2.3&  0.1421\\
13451+1232 (4C12.5)&    2098$\pm$ 315&  2060$\pm$ 185&&                 105$\pm$ 11.3&           0.12174\\
13510+0442&  1559$\pm$ 94&   2534$\pm$ 203&&         10.0$\pm$ 2.2&  0.136\\
13539+2920&  1832$\pm$ 128&  2729$\pm$ 191&&         8.5 $\pm$2.5    &       0.10845\\
14060+2919&  1611$\pm$ 160&  2417$\pm$ 170&&         9.8 $\pm$2.2    &       0.11678\\
14394+5332$^e$&  1954$\pm$ 78&   2395$\pm$ 120&  60.2$\pm$ 22& 6.5 $\pm$2.3    &       0.105\\
15001+1432$^e$&  1871$\pm$ 94&   2043$\pm$ 184&  49.0$\pm$ 12.3&  7.6 $\pm$1.9    &       0.16275\\
15206+3342&  1767$\pm$ 88&   1887$\pm$ 151&&                 3.7 $\pm$1.6$^{**}$    &       0.1244\\
15250+3609&     7286$\pm$ 364&  5907$\pm$ 295&  252.0$\pm$86$^2$&33.0 $\pm$8.0$^2$&0.055155\\
15327+2340 (Arp220)&103800$\pm$ 3390&112400$\pm$ 3372&6286.0$\pm$1482$^2$&    832.0 $\pm$86.0$^2$&    0.018\\
16474+3430&  2272$\pm$ 114&  2880$\pm$ 202&&                 11.7 $\pm$2.7   &0.11115\\
16487+5447&  2881$\pm$ 115&  3074$\pm$ 184&&&                                0.1038\\
17028+5817$^e$&  2428$\pm$ 146&  3914$\pm$ 196&&         11.2 $\pm$4.9    &       0.1061\\
\end{tabular}
\caption{ULIRG Sample and Results of Observations. 
Fluxes for the ULIRG sample from the current observations and from IRAS, including redshift data. Where no flux is given,
the object was not observed at that wavelength. Data at 60 and 100 $\mu$m from IRAS Point Source Catalog via NED.
Other data from the current observations except $^1$Fox (2000) (thesis), $^2$Dunne \& Eales, (2001), $^3$ NED, $^4$
Rigopoulou et al. (1996) (where 850 $\mu$m points are actually fluxes at 800$\mu$m), Fluxes include calibration errors assumed to be 20\% at 450$\mu$m and 10\% at 850$\mu$m added in quadrature to the observational noise.$^*$source not detected.  $^{**}$source detected at 2.3$\sigma$ significance only. Sources marked $^e$ have a weaker second component that would have fallen outside our SCUBA single point observations. Their fluxes and flux errors have been corrected for this in the manner described in the text.}

\end{table*}

\begin{table*}
\label{table3}
\begin{tabular}{cccccc}\hline
Source&F60 (mJy)&F100 (mJy)&F450 (mJy)& F850 (mJy)&   Redshift\\ \hline
09320+6134 (UGC 5101)$^2$&11680$\pm$$\pm$34&19910$\pm$137&1433$\pm$418&143$\pm$25&0.039\\
10565+2443$^1$&12100$\pm$25&15010$\pm$122&533$\pm$112&56$\pm$10&0.043\\
14348-1447$^3$&64600$\pm$38&73100$\pm$131&210$\pm$63&24$\pm$7.2&0.0827\\
16504+0228 (NGC6240)$^3$&22940$\pm$54&26490$\pm$174&1000$\pm$300&150$\pm$45&0.0244\\
17208-0014$^3$&32130$\pm$57&36080$\pm$555&1070$\pm$300&155$\pm$48&0.0428\\
22491-1808$^3$&5540$\pm$36&4640$\pm$95&&19$\pm$5.7&0.0778\\
23365+3604$^3$&7440$\pm$46&9010$\pm$216&170$\pm$51&20$\pm$6&0.0645\\
\end{tabular}
\caption{Additional Sources. 
Sources from the literature that are not part of the complete sample. $^1$ Fox, thesis; $^2$Rigopoulou et al., (1996);$^3$ Klaas et al., 2001; }
\end{table*}

%Contribution of CO to 850$\mu$m fluxes?

\section{The Dust Properties of ULIRGs}

\subsection{Dust Spectral Energy Distributions}

We parameterize the spectral energy distribution (SED) of our target ULIRGs using the standard modified black body SED, ie.

\begin{equation}
F_{\nu} = \nu^{\beta} B_{\nu}(\nu, T)
\end{equation}
The temperature and emissivity parameters for each object can be calculated using a $\chi^2$ minimization method with model and observed fluxes normalized at 60$\mu$m and T and $\beta$ as free parameters. The results of these fits are shown in Table 4.

For those objects where additional 450$\mu$m fluxes are available we also attempt a two component dust fit. This consists of dust at two different temperatures, but with a fixed emissivity parameter of $\beta$=2. This choice of $\beta$ for the two component fits is based on results from Dunne \& Eales (2001) and is discussed further in Section 4.2.

\begin{equation}
F_{\nu} = N_w \nu^{\beta} B_{\nu}(\nu, T_w) + N_c \nu^{\beta} B_{\nu}(\nu, T_c)
\end{equation}

where N$_w$ and N$_c$ represent the relative masses of the warm and cold dust components and T$_c$ and T$_w$ are their temperatures. For these fits we again normalize the fluxes at 60$\mu$m, with the three parameters, T$_c$, T$_w$ and $N_c/N_w$ calculated by $\chi^2$ minimization. An additional constraint on T$_w$ comes from the IRAS 25$\mu$m flux for these objects, which the dust emission was not allowed to exceed. The details of these fits are given in Table 5.

In both the single and two component fits we used identical methods and software to that used by the SLUGS team (Dunne et al., 2000; Dunne \& Eales, 2001; Vlahakis et al., 2005) so that our results can be directly compared to theirs. We also use the same methods for calculating the dust mass in these objects. For a single component we use:

\begin{equation}
M_d = \frac{S_{850} D^2} {\kappa_d(\nu) B(\nu, T_d)}
\end{equation}
where $S_{850}$ is the observed 850$\mu$m flux, $D$ is the luminosity distance to the object and $\kappa_d(\nu)$ is the dust opacity coefficent. We assume a value for $\kappa_d(\nu)$ of 0.077 m$^2$kg$^{-1}$, matching that of Dunne et al. (2000). For a two component dust model the mass is calculated using:
\begin{equation}
M_d = \frac{S_{850} D^2} {\kappa_d(\nu)} \times \left[ \frac{N_c}{B(\nu, T_c)} + \frac{N_w}{B(\nu,T_w)} \right]
\end{equation}
where $N_c$ and $N_w$ represent the relative mass of the cold and warm components, $T_c$ and $T_w$ are the respective temperatures for the cold and warm components, and other symbols are similar to equation 3.

\begin{table*}
\label{table4}
\begin{tabular}{ccccccc}\hline
Object&Dust Temp&Emissivity ($\beta$)&$\chi^2$&log(M$_d (M_{\odot})$)&log(L$_{850} Wm^{-2}Hz^{-1}sr^{-1})$&log(L$_{FIR} (L_{\odot}$) \\ \hline
08572+3915&64&1.2&0.005&7.38&21.92&12.03\\
10378+1108&53&1.6&0.062&7.52& 21.97& 12.25\\
10494+4424&35&2.0& 0.49&  8.02&  22.23&   12.09\\
11119+3257&51&1.5&  0.017&   7.87&  22.31&   12.41\\
11506+1331&36&2.4&  1.72&    7.67&  21.87&   12.23\\
12112+0305&43&1.4&  0.87&    8.22&  22.56&   12.23\\
12540+5708&42&1.8&  0.85&    8.09&  22.40&   12.26\\
13428+5608&43&1.6&  0.38&    7.93&  22.25&   12.01\\
13442+2321&42&1.4&  0.20&    8.22&  22.56&   12.18\\
13510+0442&37&1.8&      0.136&   8.07&  22.32&   12.14\\
13539+2920&37&1.9&  0.024&   7.84&  22.08&   11.96\\
14060+2920&39&1.7&  0.13&    7.93&  22.20&   11.99\\
14394+5332&39&1.9&  0.21&    7.67&  21.94&   11.93\\
15001+1432&47&1.6&      0.55&    7.94&  22.37&   12.34\\
15206+3342&41&2.0&  0.11&    7.52&  21.81&   12.03\\
15250+3609&55&1.1&  0.37&    7.70&  22.16&   11.91\\
15327+2340&46&1.1&  2.09&    8.28&  22.64&   12.06\\
16474+3430&42&1.6&  0.10&    7.93&  22.25&   12.06\\
17028+5817&35&2.0&  0.05&   7.97&  22.18&   12.08\\ \hline
09320+6134&  36&    1.4&  3.11&    8.28&  22.51&   11.85\\
10565+2443&  40&    1.5&  5.13&    7.90&  22.18&   11.90\\
14348-1447&  40&    1.9&  1.96&    8.04&  22.32&   12.20\\
16504+0228&  39&    1.7&  3.77&    7.88&  22.14&   11.65\\
17208-0014&  41&    1.6&  0.33&    8.32&  22.62&   12.30\\
22491-1808&  51&    1.4&  0.08&    7.77&  22.19&   12.09\\
23365+3604&  37&    2.1&  1.18&    7.82&  22.05&   12.03\\
\end{tabular}
\caption{One Component Dust Fits. 
Fits made to SCUBA 850$\mu$m fluxes, IRAS 60 and 100$\mu$m fluxes and 450$\mu$m fluxes where available. Sources above the line are part of the complete sample, those below are additional sources from the literature.}
\end{table*} 

\begin{table*}
\label{table5}
\begin{tabular}{ccccccc}\hline
Source&Warm dust T$_{H}$&Cool dust T$_C$&N$_c$/N$_w$&$\chi^2$&log(M$_d (M_{\odot}$))&log(L$_{fir} (L_{\odot}$)\\ \hline
10494+4424&     36&       33&        0.5&      0.36&    8.02&     12.09\\
%11506+1331&     36&       35&        1.1&      1.62&    7.68&     12.23\\
12112+0305&     50&       26&        21.3 &    0.06&    8.50&     12.30\\
12540+5708&      41 &      22&        1.0 &     1.12&    8.32&     12.27\\  
13428+5608&      54 &      29 &       17.3 &    0.08&    8.14 &    12.10\\
13442+2321&     39&       18  &      5.8  &    0.36 &   8.70 &    12.18\\
14394+5332&     45&       33&       4.5&     0.01  &  7.74 &    11.94\\
15001+1432&     43&       18  &      2.3  &    0.90&   8.42 &   12.34\\
15250+3609&     45 &      18  &      12.3 &    0.05 &   8.32 &    11.88\\
15327+2340&      38&       16 &       10.8 &    0.09 &   8.88  &   12.04\\ \hline
09320+6134&     40  &     23 &       17.6 &    0.21 &   8.53 &    11.89 \\
10565+2443&     39 &      24  &      3.0  &    0.03  &  8.14 &    11.90\\
14348-1447&     56   &    31  &      23.3  &   0.08 &   8.18  &   12.30 \\
16504+0228 &   52 &      26 &       30.7  &   0.54 &   8.10 &    11.76 \\
17208-0014&     39    &   21 &       2.7 &     0.47 &   8.64 &    12.30\\
23365+3604&     38&       35 &       0.3 &     0.28 &   7.81 &    12.04 \\

\end{tabular}
\caption{Two Component Dust Fits. 
Fits made to IRAS 60 and 100$\mu$m fluxes and SCUBA 450 and 850$\mu$m fluxes for a two component $\beta$=2 dust model. This fit was attempted for all sources where 450$\mu$m data was available. The 850$\mu$m luminosity is the same as that given for the object in Table \ref{table4}. The line indicates the break between those objects in the complete sample (above) and the additional literature sources (below).}
\end{table*}

\begin{figure}
\label{fig1}
\epsfig{file=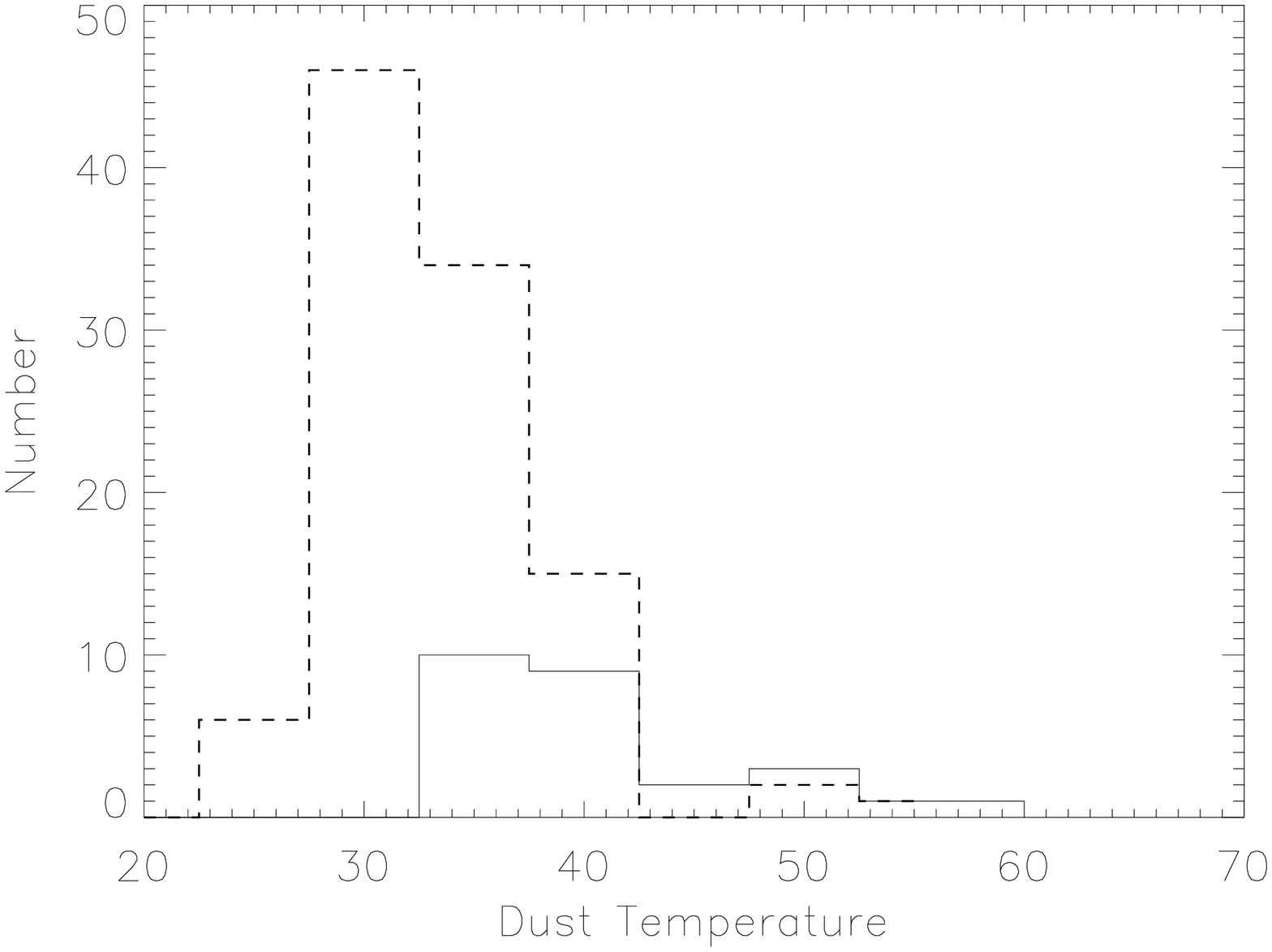,height=10cm,angle=90}
\caption{Histogram of Dust Temperatures for single temperature fits. 
ULIRGs (solid line) compared to SLUGS galaxies (Dunne et al., 2000; Vlahakis et al., 2005) (dashed line).}
\end{figure}

\begin{figure}
\epsfig{file=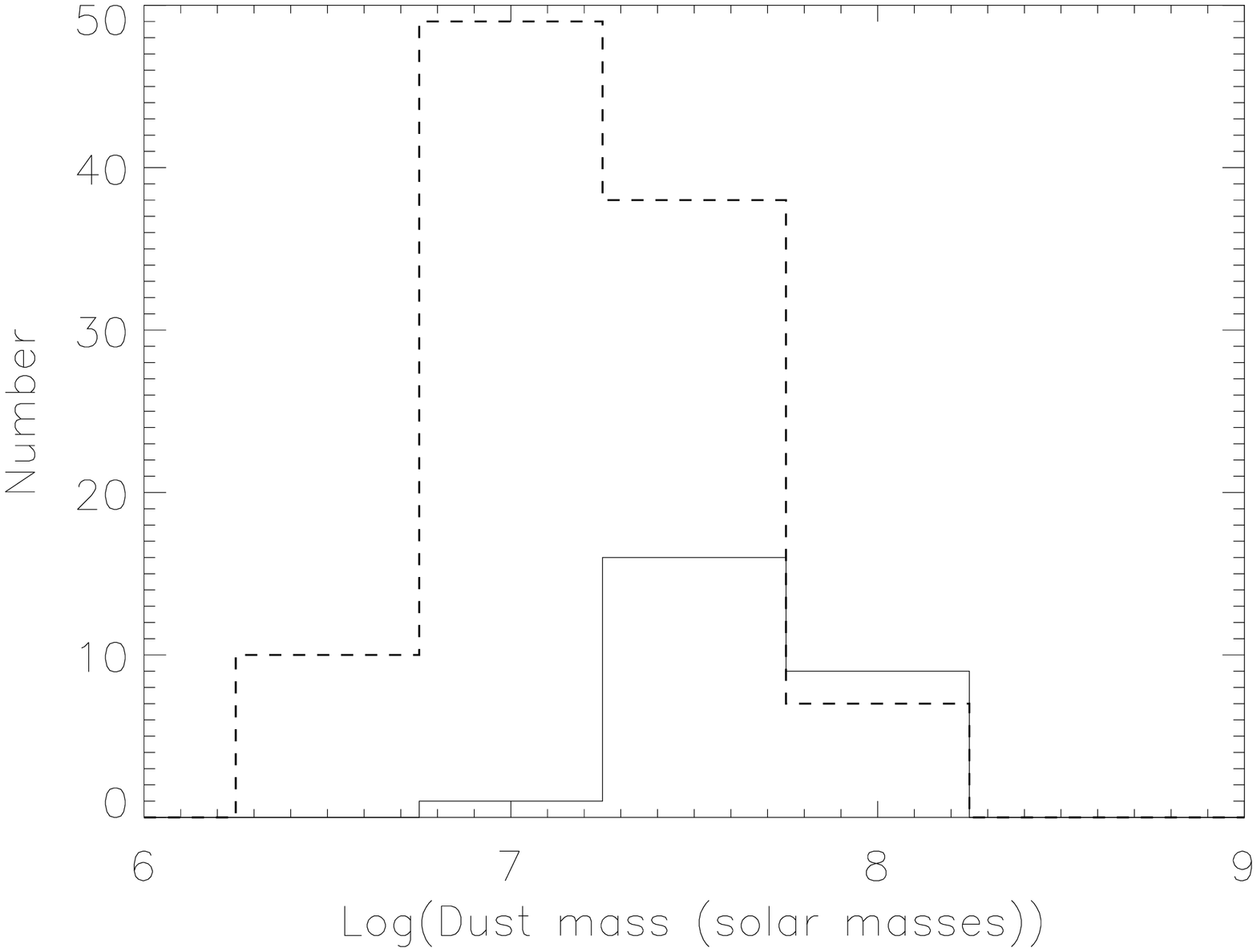,height=10cm,angle=90}
\caption{Histogram of Dust Masses for single temperature fits
ULIRGs (solid line) compared to SLUGS galaxies (Dunne et al., 2000; Vlahakis et al., 2005) (dashed line).}
\end{figure}

\begin{figure}
\label{betaplot}
\epsfig{file=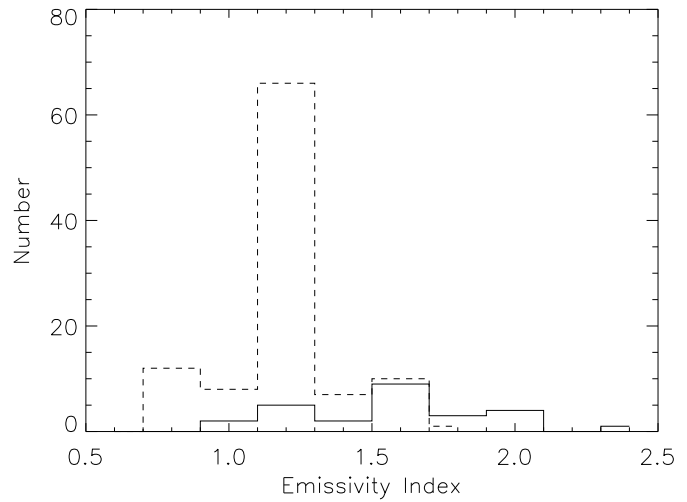,height=10cm,angle=90}
\caption{Histogram of Dust Emissivities ($\beta$) for single temperature fits.
ULIRGs (solid line) compared to SLUGS galaxies (Dunne et al., 2000; Vlahakis et al., 2005) (dashed line).}
\end{figure}

\subsection{Single Component Fits}

Figure 1 shows a comparison of the dust temperatures derived from the single component fits for our ULIRGs (complete sample and additional sources together) and for SLUGS galaxies (Dunne et al., 2000; Vlahakis et al., 2005). Figure 2 shows a similar comparison between the derived dust masses for the two sets of objects, while Figure 3 does the same for the derived emissivity index $\beta$.  The dust temperatures and emissivity index distributions for the ULIRGs and SLUGS galaxies appear to be quite different. Statistical analysis confirms this, with a KS-test showing that there is less than a 3$\times 10^{-6}$ probability that the two distributions come from the same parent distribution, with a median dust temperature of 41K compared to the SLUGS median of 35K (or means of 42.7K compared to 35.6K), and a steeper emissivity index, with a median of 1.6 compared to the SLUGS median of 1.3 (or means of 1.66 compared to 1.29). We thus conclude that ULIRGs have warmer dust and steeper $\beta$ on average than normal galaxies. In terms of the temperature-luminosity diagram of Yang et al. (2007) these single temperature fits would place our local ULIRGs in the same region occupied by their intermediate redshift ULIRGs observed with SHARC-II. This issue is discussed further in Section 4.5.

The difference in emissivities is not easy to understand in the context of a single dust component. For the emissivity to change significantly from one set of galaxies to another, there would have to be some fundamental change to the structure or makeup of the dust grains. While it is possible that such a change might occur in the harsher radiation environment of an actively star forming galaxy, such as a ULIRG, there is a much more natural explanation. This is that a single dust component is not a good model, but that we must instead consider the presence of two dust populations, cool and warm. Both sets of dust have the same emissivity, $\beta$=2, but the presence of cooler dust acts to flatten the SED at longer wavelengths, so that fits to IRAS and 850$\mu$m fluxes assuming a single dust component give a $\beta$ value lower than 2. This model can be tested when data at 450$\mu$m, or other intermediate submm wavelengths, is available and was first extensively tested by Dunne \& Eales (2001). For the fifteen ULIRGs where we have 450$\mu$m data we can make a first examination of the validity of this idea by examining the 60 $\mu$m to submm colours, as in Figure 6 of Dunne \& Eales (2001). This is shown in Figure 4. As can be seen the ULIRGs sit on exactly the same tight correlation as SLUGS galaxies, though with generally higher IRAS-to-submm ratios. They are thus likely to have similar dust properties to the SLUGS sources, albeit with higher dust temperatures. We thus attempt two component dust fits to the SEDs of these sources.

\begin{figure*}
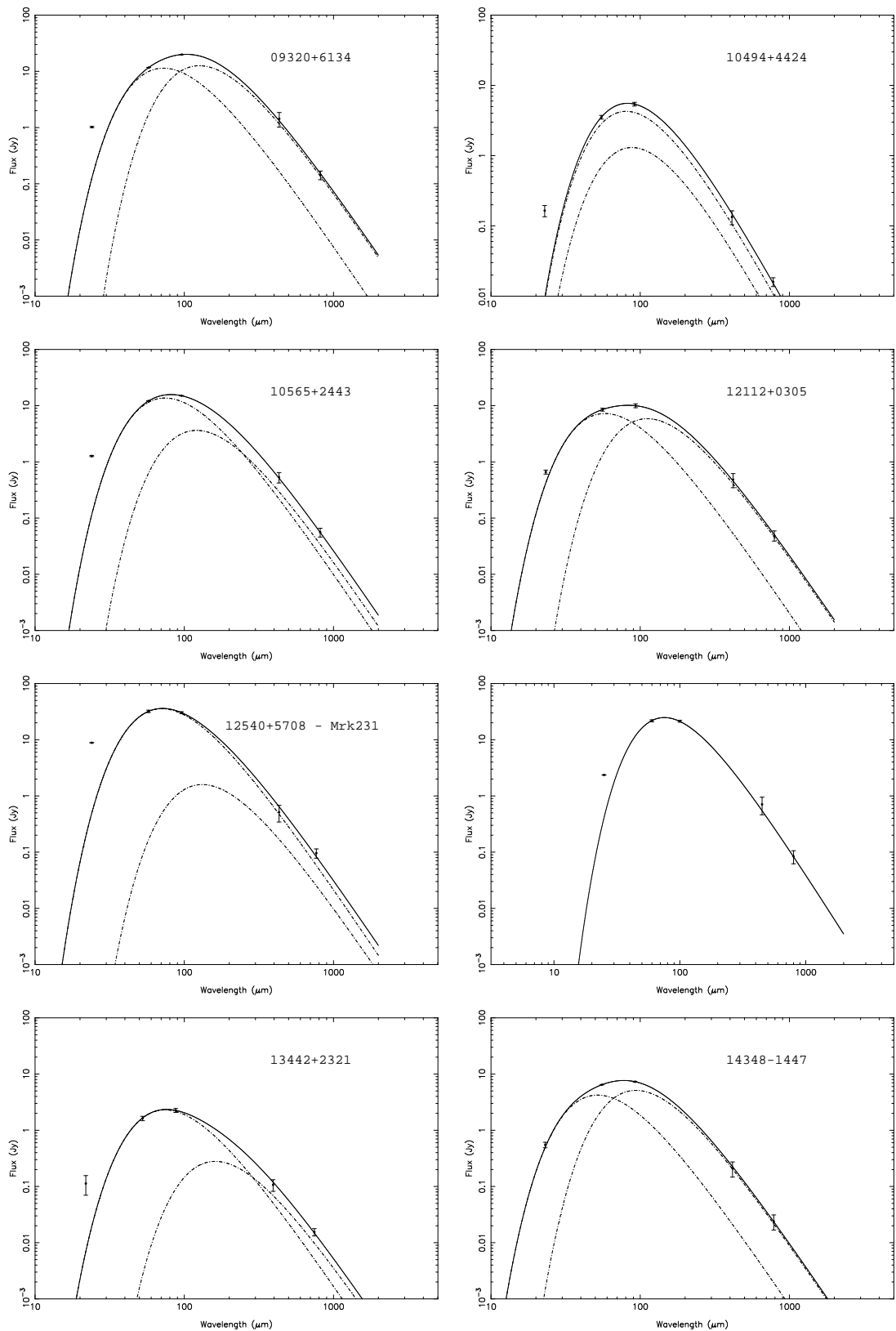

\begin{tabular}{cc}
\epsfig{file=ulirg_seds/ir09320_2.ps,height=7cm,angle=-90} &
\epsfig{file=ulirg_seds/ir10494_2.ps,height=7cm,angle=-90} \\
\epsfig{file=ulirg_seds/ir10565_2.ps,height=7cm,angle=-90} &
\epsfig{file=ulirg_seds/ir12112_2.ps,height=7cm,angle=-90} \\
\epsfig{file=ulirg_seds/mrk231_2.ps,height=7cm,angle=-90} &
\epsfig{file=ulirg_seds/mrk273.ps,height=7cm,angle=-90} \\
\epsfig{file=ulirg_seds/ir13442_2.ps,height=7cm,angle=-90} &
\epsfig{file=ulirg_seds/ir14348_2.ps,height=7cm,angle=-90} \\
\end{tabular}
\caption{Dust Spectral Energy Distribution Fits or those sources detected at both 450 and 850$\mu$m. Data points are from SCUBA (450 and 850$\mu$m) and IRAS (60 and 100$\mu$m with the IRAS 25$\mu$m flux taken as an upper limit. The solid line shows the combined SED, the dot dash lines show the warm and cool components. Sources are 09320+6134, 10494+4424, 10565+2443, 12112+0305, 12540+5708 (Mrk231), 13428+5608 (Mrk273), 13442+2321, 14348-1447, 14394+5332, 15001+1432, 15250+3609, 15327+2340 (Arp220), 17208-0014, 23365+3604, 16504+0228 (NGC6240) from left to right and then from top to bottom.}
\end{figure*}

\begin{figure*}
\begin{tabular}{cc}
\epsfig{file=ulirg_seds/ir14394_2.ps,height=7cm,angle=-90} &
\epsfig{file=ulirg_seds/ir15001_2.ps,height=7cm,angle=-90} \\
\epsfig{file=ulirg_seds/ir15250_2.ps,height=7cm,angle=-90} &
\epsfig{file=ulirg_seds/arp220_2.ps,height=7cm,angle=-90} \\
\epsfig{file=ulirg_seds/ir17208_2.ps,height=7cm,angle=-90} &
\epsfig{file=ulirg_seds/ir23365_2.ps,height=7cm,angle=-90} \\
\epsfig{file=ulirg_seds/ngc6240_2.ps,height=7cm,angle=-90}\\
\end{tabular}
\contcaption{}
\end{figure*}

\subsection{Two Component Fits}

\begin{figure}
\epsfig{file=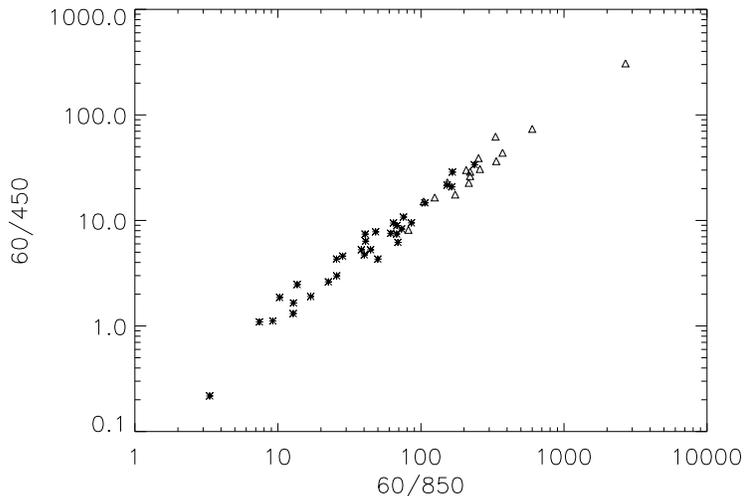,height=10cm,angle=90}
\caption{60$\mu$m to 450 and 850$\mu$m colours for ULIRGs (triangles) and SLUGS galaxies (stars) from Dunne \& Eales, 2001 and Vlahakis et al., 2005. Only those sources for which fluxes at both 850$\mu$m and 450$\mu$m are available are plotted. Note that the ULIRGs lie on the same tight correlation as the SLUGS sources suggesting that the general nature of their dust emission is very similar. The ULIRG to the extreme right of the plot is IRAS14348-1808.}
\end{figure}

\begin{figure*}
\epsfig{file=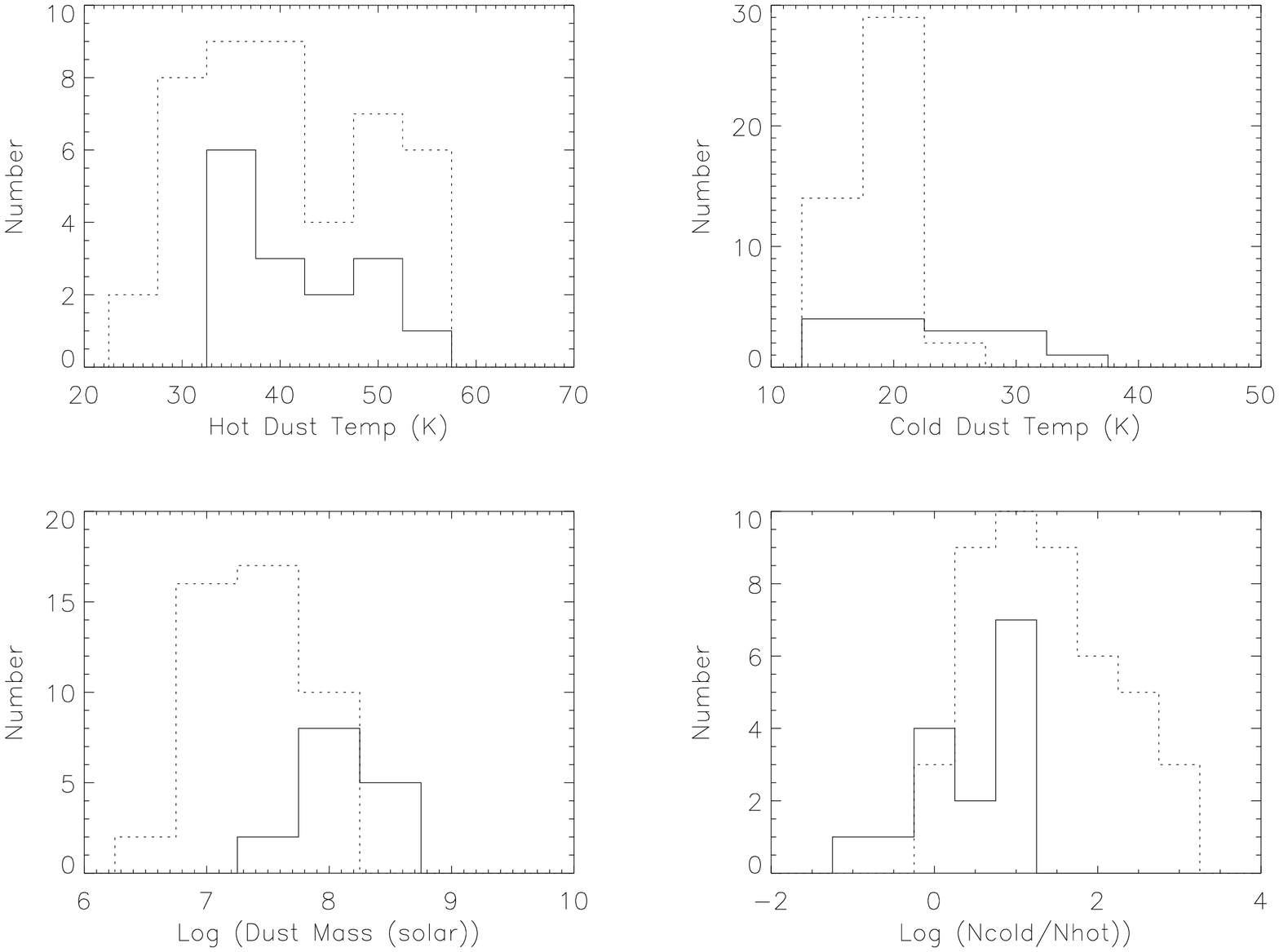,height=20cm,angle=90}
\caption{Comparison of Two Component Fits for ULIRGs and SLUGS. 
Top left: Histogram of hot dust component temperatures; Top right: Histogram of cold dust component temperatures; Bottom Left: Histogram of derived total dust masses; Bottom Right: histogram of ratios of N$_{cold}/N_{hot}$. In all cases the solid line represents the distribution of ULIRG properties and the dotted line that of the general population of dusty galaxies as found by SLUGS (Dunne \& Eales, 2001; Vlahakis et al., 2005).}
\end{figure*}

We applied two dust component fits to all fifteen objects where we have fluxes at both 450 and 850$\mu$m.  For eight of these there was clear evidence for the presence of two dust components on the basis of improved fitting (reduced $\chi^2$) compared to the one component fits. In contrast to this evidence for cold dust in 50\% of ULIRGs, nearly all normal galaxies in SLUGS where 450$\mu$m fluxes are available show the same improved fits for two dust components.

Figure 6 shows plots comparing dust temperatures, masses and the cold-to-hot ratio in the ULIRG sample and in the more general SLUGS sample. A number of things are readily apparent. Firstly the ULIRGs, as expected, seem to have significantly higher dust masses than the average SLUGS source (a KS test finds this difference significant at the 99.99\% level). Secondly the hot dust temperature distributions of the two sets of objects are similar while the cold dust temperature distributions are different (98.4\% significance) and the distribution of cold to hot dust ratios is different (99.1\% significance). ULIRGs thus seem to have less cold dust than more normal galaxies with a different temperature distribution. 

\subsection{Dust in ULIRGs}

Comparison of the two component fit results for ULIRGs and less luminous SLUGS objects suggests the following scenario. The dust content of a ULIRG is largely the combination of the dust in the precursor merger galaxies. However the merger-triggered starburst, and other associated activity, warms this dust leading, in the context of single component models, to a higher dust temperature or, in the context of two component models, to a smaller or undetectable cold dust component, as  compared to the normal SLUGS galaxies. This scenario is broadly consistent with that discussed by Klaas et al. (2001) in the context of ISOPHOT observations of ULIRGs combined with some JCMT and SEST submm data, and has also been discussed elsewhere for individual objects (eg. Clements et al. (1993) in the context of a merger-triggered LIRG).

We can test this idea further, and try to account for differences in the underlying stellar mass of the systems, by examining a plot of K-band absolute magnitude, assuming this can act as a surrogate measure of the underlying stellar mass, versus total dust mass for ULIRGs and SLUGS galaxies. This is shown in Fig. 7. As can be seen the ULIRGs and SLUGS galaxies appear to lie on the same broad correlation though the ULIRGs have more scatter about the relationship than the SLUGS objects. This will at least be partly due to AGN contributions to the K-band light, the most extreme example of which is Mrk231 which is the high luminosity outlier in the plot. Obscuration and contributions from young stars in any starburst will also boost the K-band luminosities of those objects, be they ULIRGs in our sample or LIRGs in SLUGS. In spite of this, the fact that they lie on the same broad correlation between K-band abs. mag. and dust mass suggests that the dust in ULIRGs has similar origins to that in lower luminosity systems. It is in effect the dust content of the parent galaxies of the merger combined and then heated by the starburst to higher temperatures than those found in the progenitors.

It could be argued that the somewhat simplistic model of dust adopted for these objects, with two distinct dust components at different temperatures, implies that the dust masses derived here are uncertain. While this is undoubtedly correct at some level, the fact that we are applying identical fitting procedures to similar data for ULIRGs and SLUGS galaxies nevertheless means that the relative differences between the two classes are real even if the absolute dust masses are inaccurate at some level. We can obtain a consistency check for our dust mass calculations by examining the dust-to-gas mass ratios derived from our SED fits and the CO observations of ULIRGs conducted by Solomon et al. (1997), which include ten of the ULIRGs in our sample detected at both 450 and 850 $\mu$m. The dust-to-gas mass ratios for these objects have a mean of 237 for the single temperature component fits and a mean of 101 for the two component fits. The canonical value for dust-to-gas mass ratios derived for spiral galaxies is $\sim$150 (eg. Regan et al., 2004; Bendo et al., 2006; Draine et al., 2007). This would seem to suggest that our simple dust models are doing reasonably well in estimating the mass of dust in these objects. 

Drawing conclusions about the relative masses of dust and stars in these objects is more complex. Dasyra et al. (2006) have calculated the mass of ULIRGs using stellar kinematics to measure central supermassive black hole masses and the well known correlation between these and bulge mass. Their study, including ten of the objects discussed here, finds that ULIRGs are sub-$m_*$ to $m_*$ objects. This agrees with similar calculations using emission line velocity dispersions (Colina et al., 2005). ULIRGs are nevertheless brighter than L$^*$ in the near infrared, with mean K band abs. mag. of -25.4 for our set of 15 objects with 450 and 850 $\mu$m photometry. ULIRG K-band luminosities are clearly boosted by AGN - if we remove those objects that have a clear AGN component visible in the optical, ie. those with classifications as Sy1 or Sy2, then this value drops to K = -25.0, but this is still 1.5 magnitudes above the K$^*$ value found for normal galaxies in 2MASS (Devereux et al., 2009). It thus seems that we cannot simply use K-band magnitudes as an indicator of stellar mass for this population. Further analysis of this problem must await accurate stellar mass determinations for large samples of far-IR luminous objects for which there are also dust mass determinations. The ten ULIRGs common to this paper and Dasyra et al. (2006) would make a good starting point. More broadly, Fig. 7 would seem to suggest that a broad correlation exists between K-band luminosity and dust mass for most objects. The exceptions to this correlation are those objects like Mrk 231 where an AGN is a strong contributor to the K-band light.

%f we further assume that ULRGs are the result of mergers between just two progenitor galaxies we can examine the nature of those progenitors by comparing the K-band abs. mag. and dust mass of SLUGS sources with the ULIRG values of these parameters divided by two, to provide an estimate of the progenitor properties. The results of such a calculation are shown as solid triangles in Fig. 7. On this basis we suggest that the K-band absolute magnitude of ULIRG progenitors is K=-24.5$\pm$0.2. This is about 1 magnitude brighter than the K$^*$ value found for normal galaxies in 2MASS (Devereux et al., 2009). This does not, however, mean that ULIRG progenitors are necessarily larger than L$^*$ galaxies, which would contradict the results of kinematic studies by Dasyra et al. (2006), which include ten of the same objects discussed here. Instead it suggests that we cannot simply use K-band as a simple surrogate for underlying steller mass, but must assume that the general ULIRG 

%**** remove known AGN ULIRGs and see what happens **** - remove those with Sy1 or Sy2 optical spectra

A plot of far-IR luminosity per unit dust mass against far-IR luminosity is shown in Fig. 8. This shows that ULIRGs are producing more far-IR luminosity per unit dust mass than systems with lower far-IR luminosity. This could come about either through more efficient star formation processes - i.e. more stars are produced per unit dust mass in ULIRGs - or through a tighter coupling between the dust and the sources of luminosity - i.e. a greater fraction of the luminosity produced in a ULIRG is absorbed by dust and re-readiated in the far-IR. 

%???? Add Draine Mdust Ldust points to this figure????

%We suspect this difference is a result either of absorption of some of the H-band light in the ULIRGs by dust, making them appear less luminous at this wavelength, or because we now have better determinations of the local infrared luminosity function as a result of all sky surveys such as 2MASS.

\begin{figure}
\epsfig{file=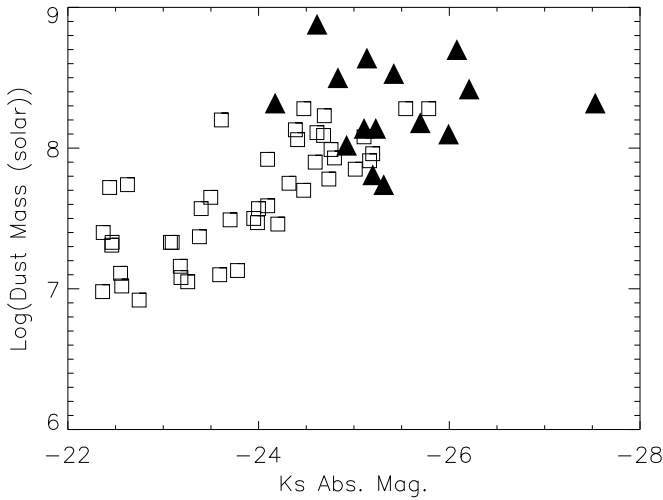,height=10cm,angle=90}
\caption{K-band absolute magnitude vs. total dust mass for ULIRGs and SLUGS galaxies. 
K-band absolute magnitude calculated using 2MASS Ks$\_$tot values and redshifts versus total dust mass derived from the two component dust fits. Open squares are SLUGS galaxies, solid triangles are ULIRGs from this paper where we have both 450 and 850$\mu$m fluxes. The outlying ULIRG with the brightest K-band absolute magnitude is Mrk231 whose K-band light will include a substantial amount of AGN emission and is thus expected to be an extreme outlier.}
\end{figure}

\begin{figure}
\epsfig{file=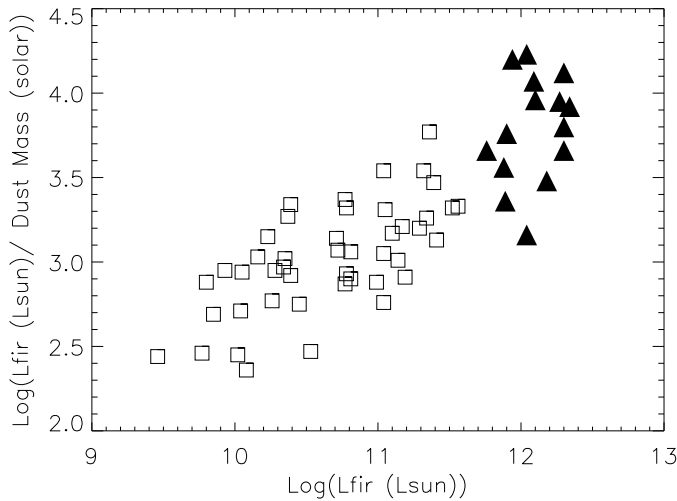,height=10cm,angle=90}
\caption{Far-IR Luminosity vs. total dust mass for ULIRGs and SLIGS galaxies.
Open squares are SLUGS galaxies, solid triangles are ULIRGs from this paper where we have both 450 and 850$\mu$m fluxes.}
\end{figure}

\subsection{Dust in SMGs}

One of the current problems in connecting local and intermediate redshift ULIRGs to the high redshift SMGs is that the two populations appear to lie in different parts of the temperature-luminosity (T-L) plane and that the SMGs appear to have higher dust masses than ULIRGs (Yang et al., 2007). In Figure 9 we plot the positions of our ULIRGs alongside SMGs from Chapman et al. (2005), Coppin et al. (2008) and Kovacs et al. (2006), as well as moderate z ULIRGs from Yang et al. (2007) and the more normal SLUGS sources from Dunne et al. (2000). As can be seen the SMGs seem to follow a rather different track in L-T space to more local objects such as ULIRGs and SLUGS galaxies, suggesting that SMGs contain systematically cooler dust than more local objects. MIPS observations of moderate redshift sources at 70 and 160$\mu$m by Symeonidis et al. (2009) appear to have found a larger population of sources with dust temperatures and luminosities similar to the SMGs, possibly representing a missing link between local and more distant far-IR luminous populations. One possible issue with these conclusions is that the dust properties of the SMGs are derived using very few photometric points - just one or two fluxes measured in the rest frame far-IR, bolstered by extrapolations from the radio to obtain an additional flux constraint using the radio-FIR relation. Even the Symeonidis et al. (2009) results are based on just two photometric points constraining the dust properties (70 and 160$\mu$m). Simple single temperature and $\beta$ dust models are all that can be fit to such data. As we have seen here with local ULIRGs such models are often an oversimplification. Only when we have observations that cover a wider range of wavelengths for SMGs will we be certain of the significance of the apparent differences between their dust properties and those of better studied, more local galaxies. 

\begin{figure}
\epsfig{file=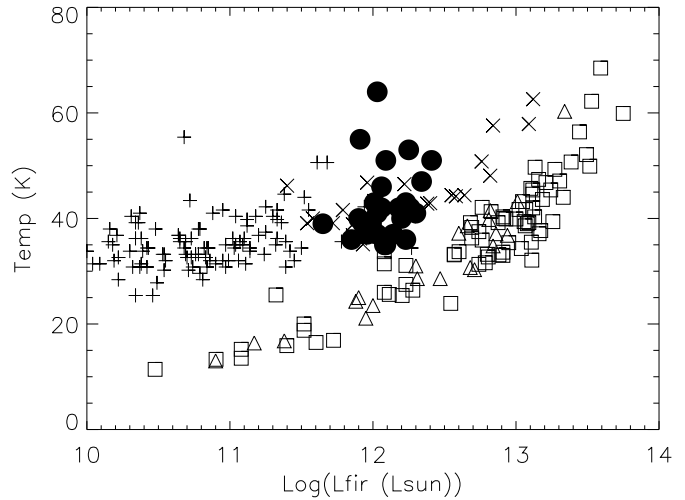,height=10cm,angle=90}
\caption{Comparison of ULIRGs, SMGs and normal galaxies on Luminosity-Temperature diagram based on single temperature dust SED fits.
Open squares are SMGs from Chapman et al. (2005), open triangles are SMGs observed with SHARCII by Coppin et al. (2006) and Kovacs et al. (2006). + signs are the SLUGS sources from Dunne et al. (2000), x are intermediate redshift ULIRGs from Yang et al. (2007), while ULIRGs from this work are solid dots.}
\end{figure}

%It is possible that there are some evolutionary trends with dust temperature throughout the evolution of a ULIRG. We investigate these by looking for differences in temperature, luminosity and dust emissivity distributions for ULIRGs known to be single or double nucleus sources, the assumption being that single nucleus sources are at a more advanced stage of merger than the double nucleus sources. We find no significant differences, but there are only 10 single and 16 double sources so our current sample size is probably too small for a definitive investigation. We note that Klass et al. (2001) tried this with their sample of 41 ISO-studied ULIRGs, but also found no evidence for such evolution.

\section{The Submillimetre Luminosity Function}

\subsection{Calculating the Luminosity Function}

Determination of the local submm luminosity function is important for comparison to the luminosity function found at high redshift for submm sources in deep fields. The current best determination of the local 850$\mu$m luminosity function comes from SLUGS (Dunne et al., 2000; Vlahakis et al., 2005). However, both of these studies include very few high submm luminosity sources. We thus use the complete ULIRG sample, i.e. those sources in Table 1, to examine the local submm luminosity function.

The luminosity function $\Phi(L) \Delta L$, giving the number density of sources (Mpc$^{-3}$) within the luminosity range $L$ to $L+\Delta L$, is given by:
\begin{equation}
\Phi(L) \Delta L = \sum_{i} 1/V_{i}
\end{equation}
where the sum is over all sources in the sample and $V_i$ is the accessible volume of the $i$th source (Avni \& Bahcall, 1980).  

We use an accessible volume method to calculate the accessible volume of a given source, the acccessible volume being the volume of space within which the source would be detected before it drops out of the sample as a result of being
%too faint at 850$\mu$m to be detected in our SCUBA observations or
too faint at 60$\mu$m to be in the initial sample. For the purposes of this calculation a 60$\mu$m flux limit of 1.5 Jy is used. K-corrections assuming  a flat SED at 60$\mu$m are used in this calculation, though the redshifts involved are sufficiently small that they make only a small difference. The resulting luminosity function is plotted in Fig. 10.

We have good 850$\mu$m fluxes for 86\% of this complete sample ie. 18/21 ULIRGs, excluding the two nonthermal objects. Of the three without good detections $>$ 3$\sigma$ one was not observed, one has a marginal detection (2.3 $\sigma$) and the last was not detected at all (0.0 $\pm$1.4 mJy flux). The two observed but undetected sources are among the faintest of our sample in the IRAS bands, so there is no evidence that they have significantly different IRAS-to-submm colours than the rest of our targets. We deal with these sources for purposes of the LF calculation in the following way. The marginally detected source has its measured flux used in spite of the low significance. The non-detected source would have a luminosity at 850$\mu$m fainter than the faintest populated ULIRG bin in Fig. 10 assuming that it has a true flux equal to the one $\sigma$ noise in the detection. It can thus be safely ignored in our discussion of the high end of the 850$\mu$m luminosity function. The non-observed source is accounted for by scaling the ULIRG luminosity values by 21/20 to account for this incompleteness factor. The two nonthermal sources have luminosities at 850$\mu$m well above those of the thermal ULIRGs plotted in Fig. 10. We will not discuss them further.

\subsection{The High End of the local 850$\mu$m Luminosity Function}

The resulting 850$\mu$m luminosity function for ULIRGs is plotted alongside the SLUGS 850$\mu$m luminosity function in Figure 10. We divide the ULIRGs up into three luminosity bins for comparison purposes. As can be seen the ULIRG LF is significantly below the SLUGS LF in its first two bins, and marginally below the SLUGS LF in the highest luminosity bin. Consideration of the SLUGS catalog explains this seeming disagreement. This shows that in SLUGS the galaxies that contribute to the highest 850$\mu$m luminosity bins are not just ULIRGs. Specifically, the highest luminosity SLUGS bin contains four objects: Arp220 and IRAS12112+0305, which are both ULIRGs, and NGC5257/8 and UGC9618 which are not. The two non-ULIRG galaxies have L$_{850} >10^{22.5} Wm^{-2} sr^{-1}$ but have L$_{FIR}$ of 10$^{11.24}$ and 10$^{11.45} L_{\odot}$ respectively ie. they are LIRGs not ULIRGs, and not even particularly luminous LIRGs. Nevertheless they are able to outshine ULIRGs at 850$\mu$m as a result of their large mass of cool dust (T=33.2K and 32K respectively for single temperature fits compared to the ULIRG mean of 41.7K, and cold dust masses of 10$^{8.44}$ and 10$^{8.65} M_{\odot}$ compared to the ULIRG mean of 10$^{8.0} M_{\odot}$). It is also worth noting that they are both close pairs of galaxies with separations large enough that they can be resolved by SCUBA. They might thus be early-stage interactions where there is as yet insufficient star formation for their dust to be warmed to average ULIRG temperatures and for their far-IR luminosities to become comparable to ULIRGs.

Clearly our ULIRG sample does not include all of the most luminous objects at 850$\mu$m so our measurement of the high end of the 850$\mu$m luminosity function is an underestimate. The relative numbers of ULIRGs and non-ULIRGs in the highest luminosity bin can be estimated from the comparison to SLUGS which suggests that ULIRGs contribute only 50\% of the objects in this luminosity bin. An estimate of the true luminosity function value in this final bin, including both ULIRGs and non-ULIRGs, can be found by multiplying our ULIRG-only value by 2. This corrected value is also shown in Figure 10 and is consistent with the LF found by SLUGS, though the error bars remain large because of the large correction and high uncertainty in the true value of the correction. This would seem to confirm the earlier result of SLUGS that the 850$\mu$m luminosity function has a different shape to that of simply-extrapolated 60$\mu$m luminosity functions. 

At high luminosities the SLUGS survey itself is potentially incomplete because all the high luminosity sources in SLUGS are selected from the IRAS Bright Galaxy Sample and thus require a detection in the IRAS 60$\mu$m band. The optically selected SLUGS extension (Vlahakis et al., 2005) demonstrated the presence of cooler dust in many galaxies below the BGS 60$\mu$m selection threshold with temperatures lower than 30K in many cases (for single temperature fits). If such objects can arise with high 850$\mu$m luminosities and if none of them were sampled in the SLUGS IRAS selected survey, then the high end of the local 850$\mu$m luminosity function plotted in Figure 10 could still be a lower limit. Such high luminosity cold (T$<$30K) dust systems, have been suggested in recent followup work to the SWIRE survey (Rowan-Robinson et al., 2008; Clements et al., 2008), and in observations of SN1a host galaxies at z$\sim$0.5 (Clements et al., 2005; Farrah et al., 2004).  We conclude that the best way to make progress in determinations of the local 850$\mu$m luminosity function are large area unbiased surveys. Such observations will become possible with the new generation of submm instruments such as SCUBA2. One such survey, the SCUBA2 All Sky Survey (SASSy), is already planned (Thompson et al., in preparation). Meanwhile, the Herschel-ATLAS survey will provide a large area ($\sim$550 sq. deg.) unbiased survey at five wavelength from 100 to 500 $\mu$m. This will provide much better coverage of the far-IR/submm SED, allowing better determination of dust temperatures and SEDs, as well as detecting $\sim$ 10$^5$ galaxies out to z$\sim$0.3 (Eales et al., 2009). This project will finally determine the role of cold dust in local galaxies and thus allow much clearer insights into the nature of the far-IR population both locally and at higher redshift.

\begin{figure}
\epsfig{file=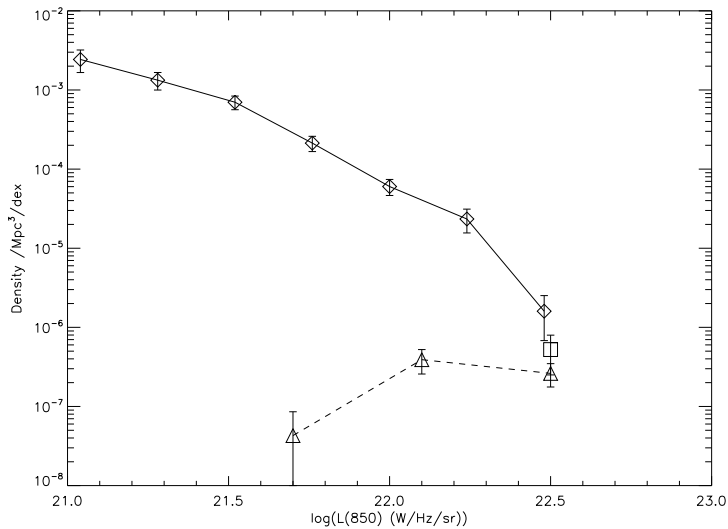,height=10cm,angle=90}
\caption{The 850$\mu$m Luminosity Function from ULIRGs and SLUGS. 
Diamonds and the solid line show the SLUGS luminosity function for 850$\mu$m sources, triangles and the dashed line show the 850$\mu$m luminosity function for ULIRGs as discussed here. The square shows the highest luminosity bin for the ULIRGs multiplied by two, as described in the text, to account for the non-ULIRG fraction of sources in this luminosity bin found by SLUGS.}
\end{figure}

\section{Conclusions}

We have obtained submm photometry at 850$\mu$m for a nearly complete sample of local ULIRGs. We use this data, combined with similar observations of a number of objects in the literature, to examine
the dust properties of ULIRGs. Dust SEDs are fit using the same techniques as those applied to the SLUGS survey (Dunne et al., 2000; Vlahakis et al., 2005) and we compare our results with these surveys' results on lower luminosity systems. We find that ULIRGs have higher dust temperatures (42K compared to 35K), higher dust masses and somewhat higher emissivity indexes. 
The higher emissivity index is explained  in the context of two component dust models, one cold and one hot. We investigate this using two component dust fits to those ULIRGs where we have 450$\mu$m as well as 850$\mu$m data. We find evidence to support the idea that ULIRGs have less cold dust than more normal galaxies leading to a steepening of the dust SED and, when this is fit by a single temperature model, a higher emissivity index. We then examine the relationship between dust mass and far-IR luminosity in ULIRGs and the lower luminosity SLUGS sources. We conclude that the dust mass of ULIRGs is the combination of the contents of the galaxies whose merger is driving the ULIRG activity. The cool dust in the progenitors is heated to higher temperatures by the ULIRG's starburst accounting for the different ratio of cool to warm dust seen in ULIRGs and lower luminosity systems. Comparison of ULIRG properties with distant SMGs suggests that the cold dust found in our two component dust models is similar to the dust component seen in SMGs at emitted-frame wavelengths of about 250$\mu$m. The suggested dichotomy between local ULIRGs and SMGs in the L-T plane may simply be the result of observational constraints and not an inherent difference between the two classes of object.

We also use our complete sample to examine the high end of the local 850$\mu$m luminosity function. We find that this is consistent with the luminosity function found in SLUGS but that a full assessment is complicated by the fact that roughly 50\% of  sources in the highest 850$\mu$m luminosity bin are not in fact ULIRGs.

This study presents the largest SCUBA dataset on local ULIRGs currently available. The next few years will see considerable advances beyond what has been achieved with SCUBA. SCUBA2 will allow deeper 850 and 450$\mu$m observations than SCUBA and will also allow us to make unbiased surveys of submm sources through large area blank field surveys such as SASSy (Thompson et al., in preparation). New generations of 350$\mu$m instruments are already here, but have yet to be extensively deployed in observations of ULIRGs. Nevertheless, some interesting results have already been achieved (Khan et al., 2005; Yang et al., 2007). The Herschel Space Telescope, operating out to 500$\mu$m in wavelength, will provide unprecedented new insights into the cold dust content of galaxies. The data presented here will be useful for all these future projects as it compiles what has been achieved on ULIRGs with SCUBA, the highly successful first generation submm bolometer array.
\\~\\
{\bf Acknowledgements}
The James Clerk Maxwell Telescope is operated by The Joint Astronomy Centre on
behalf of the Particle Physics and Astronomy Research Council of the United Kingdom,
the Netherlands Organisation for Scientific Research, and the National Research Council of Canada. DLC and LD acknowledge funding from PPARC and STFC. This research has made use of the NASA/IPAC Extragalactic Database (NED) and the NASA/ IPAC Infrared Science Archive which are operated by the Jet Propulsion Laboratory, California Institute of Technology, under contract with the National Aeronautics and Space Administration. We thank George Bendo, Katherine Vlahakis and Pierre Chanial for helpful discussions, and the anonymous referee for suggestions that have significantly improved the paper.
\\~\\

\end{document}